\documentclass[aip,sd,preprint,numerical,amsmath,amssymb,floatfix]{revtex4-2}

\usepackage[T1]{fontenc}
\usepackage[utf8]{inputenc}
\usepackage{graphicx}
\usepackage{amsmath}
\usepackage{mathtools}
\usepackage{bm}
\usepackage[
  colorlinks=true,
  urlcolor=black,
  linkcolor=black,
  citecolor=black
]{hyperref}
\newcommand{\mgcl}{MgCl$_2$}
\newcommand{\nasulf}{Na$_2$SO$_4$}
\newcommand{\mgso}{MgSO$_4$}
\newcommand{\Deltaalpha}{\Delta\alpha}

\begin{document}

\title{Terahertz-induced local-field dynamics and transient birefringence in aqueous electrolytes}

\author{Naveen Kumar Kaliannan}
\affiliation{Dynamics of Condensed Matter and Center for Sustainable Systems Design, Chair of Theoretical Chemistry, University of Paderborn, Warburger Str. 100, 33098 Paderborn, Germany}

\author{Thomas D. K{\"u}hne}
\email{tkuehne@cp2k.org}
\affiliation{Center for Advanced Systems Understanding (CASUS), Conrad-Schiedt-Stra{\ss}e 20, 02826 G{\"o}rlitz}
\affiliation{Helmholtz-Zentrum Dresden-Rossendorf, Bautzner Landstra{\ss}e 400, 01328 Dresden, Germany}
\affiliation{Institute of Artificial Intelligence, Technische Universit\"at Dresden, Helmholtzstra{\ss}e 10, 01069 Dresden, Germany}

\author{Hossam Elgabarty}
\affiliation{Department of Chemistry, University of Paderborn, Warburger Str. 100, 33098 Paderborn, Germany}

\date{\today}

\begin{abstract}

Terahertz Kerr-effect (TKE) spectroscopy provides a time-domain optical probe of the intermolecular structural dynamics of liquids, but the measured birefringence can only be interpreted microscopically if transient molecular structure, local electric fields, and nonlinear optical response are treated on the same footing.
We combine terahertz-driven non-equilibrium molecular dynamics with a hyperpolarizability-enhanced dipole-induced-dipole (DID) response model to connect the measured TKE signal of water and aqueous electrolytes to molecular-scale hydration dynamics.
Starting from a multipolar expansion of the field-dependent molecular energy, the optical response is written as a differential polarizability in which first and second molecular hyperpolarizabilities are coupled to the instantaneous local field generated by the evolving liquid structure.
For neat water, these local-field terms convert an insufficiently structured intramolecular response into the experimentally observed bipolar TKE line shape.
For aqueous \mgcl{}, polarizable force-field trajectories combined with the extended optical response reproduce the concentration-dependent enhancement of the negative birefringence, whereas non-polarizable trajectories yield the wrong trend.
Species-resolved analysis assigns the high-concentration response to strongly field-polarized anions and, more importantly, to water molecules bridging cations and anions.
The work identifies TKE birefringence as a sensitive readout of ion-hydration local-field dynamics and shows that induced polarization in both the trajectory and the optical response is required to recover the structural dynamics encoded in electrolyte TKE measurements.
\end{abstract}

\maketitle

\section{Introduction}

Time-resolved structural science increasingly relies on observables that do not image nuclear coordinates directly, but encode the evolving arrangement of molecules through fields, response functions, and ultrafast relaxation pathways.
Liquids and electrolyte solutions are a particularly important case: their ``structure'' is an ensemble of fluctuating solvation motifs, hydrogen-bond configurations, and local electric fields, and their dynamics unfold through coupled orientational, translational, and electronic-polarization degrees of freedom.
The central problem is therefore to connect an experimentally accessible ultrafast signal to the molecular sub-ensembles and local fields that generate it.

The optical response of a condensed phase is governed by the polarization induced by external electromagnetic fields.
In linear spectroscopy, this response is described by the first-order susceptibility, whereas nonlinear optical measurements access higher-order polarization components and thereby provide information on intermolecular structure, orientational dynamics, and local-field fluctuations \cite{Mukamel1995}.
Among nonlinear techniques, optical Kerr-effect and terahertz Kerr-effect spectroscopy are particularly sensitive to low-frequency collective motions because they monitor the transient birefringence induced by an impulsive electric field \cite{Bursulaya1997,Kampfrath2018,kampfrath22018,Zhao2020,Zhao2021}.

Terahertz Kerr-effect (TKE) spectroscopy is especially well suited for aqueous systems.
The driving terahertz pulse couples to rotational, librational, and translational degrees of freedom rather than to intramolecular covalent stretches, while the optical probe detects the induced anisotropy of the refractive index.
In this sense, TKE is a structural-dynamics observable: the signal follows how the liquid reorganizes after an impulsive field perturbation, while the optical readout weights those reorganizations by the instantaneous molecular polarizability and local field.
Recent experiments have shown that neat water and concentrated aqueous salt solutions display markedly different TKE signals \cite{HuangHaoyu2020,Hossamenergytransfer,Balos2022}.
In neat water, the response has a characteristic bipolar shape associated with intermolecular hydrogen-bond motion and relaxation of excess translational kinetic energy.
In concentrated electrolytes such as \mgcl{}, the negative birefringence is strongly enhanced and depends sensitively on salt concentration and ion identity.

Interpreting these experiments requires a molecular expression for the time-dependent polarizability anisotropy.
The refractive index of a liquid is related to its polarizability through Lorentz--Lorenz and Clausius--Mossotti-type relations \cite{claussismossotirelation1932}.
In a molecular simulation, however, the relevant quantity is not the gas-phase molecular polarizability alone, but the collective polarizability generated by local electric fields and intermolecular collision-induced interactions.
Dipole-induced-dipole (DID) models provide a computationally efficient route to this collective response by coupling the polarizabilities of neighboring molecules through the dipole--dipole interaction tensor \cite{Thole1981,Duijnen1998,torri2000,zhangDID,polok2018}.
The accuracy of such models can be improved by anisotropic molecular polarizabilities, damping functions, dipole--quadrupole corrections, distance-dependent polarizabilities, and hyperpolarizability terms \cite{PeterHammHuiszoon,Torri2002,skaf2003,skaf2005,Elola2007,tang2018,Harczuck2016,skafPRL2005}.
Hyperpolarizability is not only a gas-phase molecular property: in water it is modulated by solvent fluctuations and nuclear motion \cite{Liang2017}.
This makes local-field coupling especially relevant for hydrogen-bonded and ionic liquids, where the electric field experienced by a molecule is heterogeneous and time dependent.

A second requirement is an accurate molecular trajectory.
\textit{Ab initio} molecular dynamics can describe electronic polarization and changes in the local electronic structure on the fly \cite{MarxHutter2009,thomaswire2014,Hutter2014WIREsCP2K,HutterIannuzziKuehne2023,thomascp2k2020,cp2k2026madesimple}, but it remains expensive for the system sizes and trajectory ensembles required to converge weak nonlinear signals.
Classical force-field molecular dynamics is far less expensive and can produce the large statistical ensembles needed for TKE simulations, but its predictive power depends on the quality of the force field.
For aqueous electrolytes, induced polarization is expected to be important because ion--water interactions generate strong and spatially heterogeneous local fields.

Here we combine non-equilibrium molecular dynamics (NEMD) simulations with a hyperpolarizability-enhanced DID model to simulate the TKE-induced polarizability anisotropy of pure water and aqueous ionic solutions.

The central objective is to turn the TKE observable into a molecularly resolved probe of electrolyte structural dynamics.
First, we derive a compact working expression for the classical and hyperpolarizability-extended DID response used in the simulations, with the detailed multipolar derivation given in the supplementary material.
Second, we assess the importance of first and second hyperpolarizability terms by comparing computed signals with available TKE measurements.
Third, we decompose the response into ionic and water sub-ensembles to identify the molecular origin of the enhanced birefringence in concentrated salt solutions.
This decomposition is the key structural step: it identifies which hydration motifs, rather than which chemical species alone, dominate the transient optical anisotropy.

\section{Computational approach}
\label{sec:methods}

\subsection{Terahertz-driven molecular dynamics trajectories}

We performed NEMD simulations of neat water and aqueous salt solutions under an externally applied terahertz electric field.
The trajectories provide the time-dependent structural ensemble: molecular positions define hydrogen-bond and ion-hydration motifs, while the force-field polarization variables define the local electronic response sampled during the field-driven motion.
The simulated systems are summarized in Table~\ref{tab:systems}.
All trajectories were propagated in the microcanonical ensemble after canonical equilibration, using periodic boundary conditions and a 0.4~fs time step.
The applied field had the same pulse shape as in the corresponding experimental and simulation studies \cite{Hossamenergytransfer,Balos2022}; in the simulations, the amplitude was increased by a factor of eight to improve the signal-to-noise ratio while remaining in the perturbative response regime.
Trajectory frames were stored every 80 time steps.

\begin{table}
\caption{\label{tab:systems}
Simulation systems used for the TKE response calculations.
PFFMD and FFMD denote polarizable and non-polarizable force-field molecular dynamics, respectively.
The number of independent trajectories is given in brackets.
}
\begin{ruledtabular}
\begin{tabular}{lcccl}
System & Concentration & Ions & Water & Trajectories \\
 & (mol/l) &  & molecules &  \\
\hline
Water & -- & -- & 128 & PFFMD [90k], FFMD [150k] \\
\mgcl{} & 1 & 2 & 125 & PFFMD [90k], FFMD [150k] \\
\mgcl{} & 2 & 5 & 121 & PFFMD [90k], FFMD [150k] \\
\mgcl{} & 4 & 8 & 101 & PFFMD [90k], FFMD [150k] \\
\nasulf{} & 1 & 2 & 125 & PFFMD [90k], FFMD [150k] \\
NaCl & 2 & 5 & 108 & PFFMD [90k], FFMD [150k] \\
NaCl & 4 & 8 & 104 & PFFMD [90k], FFMD [150k] \\
\mgso{} & 2 & 5 & 125 & PFFMD [90k], FFMD [150k] \\
NaF & 1 & 3 & 137 & PFFMD [56k] \\
\end{tabular}
\end{ruledtabular}
\end{table}

Polarizable force-field molecular dynamics (PFFMD) simulations were carried out with TINKER \cite{tinker8}.
The AMOEBA force field with the \texttt{amoebanuc17} parameter set was used for water and common ions \cite{Ponder2010,pengyu2003,yue2013,amoebapolarisability1}.
Parameters for sulfate were taken from previous AMOEBA-based studies \cite{lambrechtso42011,florian2017}.
Short-range interactions were described by the buffered 14-7 potential \cite{halgren1992}, and long-range electrostatics were treated with particle-mesh Ewald summation \cite{Ulrich1995}.
The mutual induced dipoles were converged with a preconditioned conjugate-gradient solver to a tolerance of $10^{-5}$~D \cite{WangPCG2005,aviatPCG2017}.
The external-field force and induced-dipole terms were implemented in TINKER with unit-conversion factors appropriate for fields in atomic units and polarizabilities in \AA$^3$.

Non-polarizable force-field molecular dynamics (FFMD) simulations were performed with GROMACS 2020.4 \cite{Gromacssoft}.
Water was represented by the rigid SPC/E model \cite{berendsen1987}, and ion parameters were taken from AMBER-compatible parameter sets \cite{cornell1995,kashefolgheta2017,kashefolgheta2018}.
Long-range electrostatics were computed with particle-mesh Ewald summation \cite{Ulrich1995}; Lennard-Jones interactions were truncated using a shifted potential \cite{Lennard_Jones_1931}.
Water geometry was constrained with the LINCS algorithm \cite{BerkLINCS1997}.

Hydrogen bonds were identified using the Luzar--Chandler criterion, i.e., an O--O distance below 3.5~\AA{} and an O--H$\cdots$O angle below $30^\circ$ \cite{luzar1996}.
Ion solvation shells were defined by the first minimum of the radial distribution function between the ion and the water center of mass \cite{zhangDID}.

\subsection{Molecular response parameters for the optical readout}

The molecular polarizability model uses permanent molecular polarizabilities, induced polarizabilities, and the first and second molecular hyperpolarizabilities of water.
These quantities translate the structural snapshots into the optical observable and therefore determine how local hydration fields are weighted in the simulated TKE response.
The water molecular frame is defined with the $x$ axis along the H--H vector, the $y$ axis along the molecular bisector, and the $z$ axis perpendicular to the molecular plane.
The gas-phase water parameters used in the present calculations are listed in Table~\ref{tab:water_parameters}.
The permanent dipole moment and polarizability were parameterized with CP2K using the Gaussian and plane-wave formalism \cite{thomascp2k2020,cp2k2026madesimple}.
The first and second hyperpolarizabilities were evaluated with DALTON using quadratic and cubic response theory, respectively \cite{aidasDALTON2014,Vahtras1992,Michal1993,Jonsson1996,norman1995}.
The electronic-structure calculations used the BLYP exchange-correlation functional, Grimme D3 dispersion corrections, Goedecker--Teter--Hutter pseudopotentials, and a 400~Ry plane-wave cutoff for the charge density \cite{GrimmeD3,Morawietz8368,thomascp2k2020}.
Input files, scripts, and analysis code are available through the associated Zenodo archive and public code repository \cite{kaliannanzenodo2022,kaliannandipol2022}.

\begin{table}
\caption{\label{tab:water_parameters}
Gas-phase water dipole, polarizability, first hyperpolarizability, and second hyperpolarizability parameters.
Values in parentheses are given in atomic units.
}
\begin{ruledtabular}
\begin{tabular}{lcc}
Quantity & Value & Unit \\
\hline
$\mu_y$ & 1.93 (0.759) & D \\
$\alpha_{xx}$ & 1.3725 (9.2713) & \AA$^3$ \\
$\alpha_{yy}$ & 1.1580 (7.8224) & \AA$^3$ \\
$\alpha_{zz}$ & 0.9127 (6.1653) & \AA$^3$ \\
$\beta_{yyy}$ & $-0.5282$ ($-12.730$) & \AA$^5$ \\
$\beta_{xxy}$ & $-0.6824$ ($-16.445$) & \AA$^5$ \\
$\beta_{zzy}$ & $-0.2297$ ($-5.5365$) & \AA$^5$ \\
$\gamma_{xxxx}$ & 3.1351 (269.8054) & \AA$^7$ \\
$\gamma_{yyyy}$ & 1.4700 (126.5123) & \AA$^7$ \\
$\gamma_{zzzz}$ & 0.09407 (8.0962) & \AA$^7$ \\
$\gamma_{xxyy}$ & 2.1019 (180.8921) & \AA$^7$ \\
$\gamma_{xxzz}$ & 1.0817 (93.0972) & \AA$^7$ \\
$\gamma_{yyxx}$ & 2.1019 (180.8921) & \AA$^7$ \\
$\gamma_{yyzz}$ & 0.4124 (35.4941) & \AA$^7$ \\
$\gamma_{zzxx}$ & 1.0817 (93.0972) & \AA$^7$ \\
$\gamma_{zzyy}$ & 0.4124 (35.4941) & \AA$^7$ \\
\end{tabular}
\end{ruledtabular}
\end{table}

\subsection{Hyperpolarizability-enhanced DID readout of local-field dynamics}

The optical probe in a TKE experiment measures the field-induced change in the molecular dipole response, not simply the zero-field gas-phase polarizability.
Consequently, the simulated observable must be evaluated along the evolving nonequilibrium trajectory rather than from a static structural descriptor.
For a molecule in a local electric field $\mathbf{E}_i^{\mathrm{loc}}$, the dipole response can be written, after neglecting field-gradient terms beyond dipole order, as
\begin{equation}
\mu_{\alpha} =
\mu_{\alpha}^{0}
+ \alpha_{\alpha\beta}^{0} E_{\beta}
+ \frac{1}{2}\beta_{\alpha\beta\gamma} E_{\beta}E_{\gamma}
+ \frac{1}{6}\gamma_{\alpha\beta\gamma\delta} E_{\beta}E_{\gamma}E_{\delta}
+ \cdots ,
\label{eq:dipole_expansion}
\end{equation}
where repeated Cartesian indices imply summation.
The differential polarizability detected by the optical probe is therefore
\begin{equation}
\Pi_{\alpha\beta} =
\frac{\partial \mu_{\alpha}}{\partial E_{\beta}}
=
\alpha_{\alpha\beta}^{0}
+ \beta_{\alpha\beta\gamma} E_{\gamma}
+ \frac{1}{2}\gamma_{\alpha\beta\gamma\delta} E_{\gamma}E_{\delta}
+ \cdots .
\label{eq:differential_polarizability}
\end{equation}
Equations~\eqref{eq:dipole_expansion} and \eqref{eq:differential_polarizability} provide the formal basis for the hyperpolarizability terms used below.
The full derivation from a multipolar expansion of the interaction energy is given in the supplementary material.

The TKE signal is compared to the transient polarizability anisotropy
\begin{equation}
\Deltaalpha(t) =
\alpha_{xx}(t) - \frac{\alpha_{yy}(t)+\alpha_{zz}(t)}{2},
\label{eq:anisotropy}
\end{equation}
where the applied terahertz field is taken along the $x$ direction.
The total polarizability tensor is written as a sum over permanent and induced molecular contributions,
\begin{equation}
\bm{\alpha}_{\mathrm{tot}} =
\sum_{i=1}^{N}
\left(
\bm{\alpha}_{i}^{\mathrm{perm}} +
\bm{\alpha}_{i}^{\mathrm{ind}}
\right).
\label{eq:total_alpha}
\end{equation}
Ion permanent polarizabilities were taken from the polarizable force-field parameter sets \cite{amoebapolarisability1,amoebapolarisability2,polarisabilityso42,polarisabilityCAnMg}.
For water, the anisotropic molecular polarizability in Table~\ref{tab:water_parameters} was used.

The induced contribution was obtained by solving a self-consistent DID problem.
In its classical form, the induced tensor of molecule $i$ is
\begin{equation}
\bm{\alpha}_{i}^{\mathrm{ind}} =
\bm{\alpha}_{i}^{\mathrm{perm}}
\sum_{j\ne i}^{N}
\mathbf{T}_{ij}
\left(
\bm{\alpha}_{j}^{\mathrm{perm}}+
\bm{\alpha}_{j}^{\mathrm{ind}}
\right),
\label{eq:did}
\end{equation}
where
\begin{equation}
\mathbf{T}_{ij} =
\frac{3\mathbf{r}_{ij}\mathbf{r}_{ij}^{T}}{r_{ij}^{5}}
-\frac{\mathbf{I}}{r_{ij}^{3}}
\label{eq:tensor}
\end{equation}
is the dipole--dipole interaction tensor.
To include local-field--hyperpolarizability coupling, Eq.~\eqref{eq:did} was extended as
\begin{equation}
\bm{\alpha}_{i}^{\mathrm{ind}} =
\bm{\alpha}_{i}^{\mathrm{perm}}
\sum_{j\ne i}^{N}
\mathbf{T}_{ij}
\left(
\bm{\alpha}_{j}^{\mathrm{perm}}+
\bm{\alpha}_{j}^{\mathrm{ind}}
\right)
 + \bm{\beta}_{i}\cdot\mathbf{E}_{i}
 + \frac{1}{2}\mathbf{E}_{i}\cdot\bm{\gamma}_{i}\cdot\mathbf{E}_{i}.
\label{eq:did_extended}
\end{equation}
The local field $\mathbf{E}_{i}$ contains the field of all permanent charges and permanent and induced dipoles, but excludes the external terahertz field,
\begin{equation}
\mathbf{E}_{i} =
\sum_{j\ne i}
\frac{q_j\mathbf{r}_{ij}}{r_{ij}^{3}}
+ \sum_{j\ne i}
\mathbf{T}_{ij}
\left(\boldsymbol{\mu}_{j}^{\mathrm{perm}}+\boldsymbol{\mu}_{j}^{\mathrm{ind}}\right).
\label{eq:local_field}
\end{equation}
The permanent dipole moment of ions was set to zero, while water was assigned the gas-phase dipole moment in Table~\ref{tab:water_parameters}.
The induced dipoles entering Eq.~\eqref{eq:local_field} were obtained self-consistently with the same molecular polarizabilities and local fields used in the trajectory analysis.
The self-consistent equations were solved until the induced polarizability changed by less than $10^{-6}$~\AA$^3$.
Thole damping was applied to prevent short-range polarization instabilities \cite{Thole1981,Duijnen1998,polok2018}.
The intermolecular cutoff used for the collision-induced response was $r_t=7.5$~\AA{} unless otherwise stated.

\begin{figure}
\centering
\includegraphics[width=0.90\textwidth]{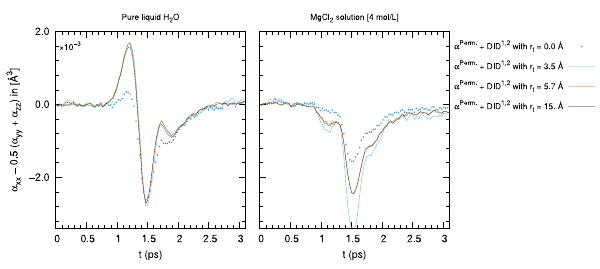}
\caption{\label{fig:cutoff}
Dependence of the computed transient polarizability anisotropy of neat water and 4~mol/l \mgcl{} on the collision-induced cutoff radius $r_t$.
The neat-water response is essentially converged already at short range, whereas the concentrated salt solution is sensitive to the local-field environment up to approximately the second solvation shell.
The cutoff dependence therefore identifies the spatial range of the hydration structure that is optically sampled by the TKE response.
}
\end{figure}

\section{Results and discussion}
\label{sec:results}

\subsection{Spatial range of the local-field structural response}

Figure~\ref{fig:cutoff} shows the effect of the intermolecular cutoff used in the DID response.
For pure water, the polarizability anisotropy is nearly insensitive to increasing $r_t$ beyond 3.5~\AA{}.
This indicates that the dominant collision-induced contribution is local and largely controlled by the first hydrogen-bond shell.
In 4~mol/l \mgcl{}, by contrast, the response changes substantially between 3.5 and about 5.7~\AA{}.
The larger cutoff sensitivity reflects the longer-ranged and more heterogeneous local electric fields generated by the ions and their solvation shells.
The response changes only weakly beyond this range, justifying the use of $r_t=7.5$~\AA{} in the production analysis.
Thus, the TKE response of concentrated electrolytes is not a purely nearest-neighbor observable.
It reports on a mesoscopic local hydration environment extending from the first ion-water contacts into the second solvation shell, precisely the spatial range over which ion pairing and shared hydration waters reorganize after terahertz excitation.

\subsection{Hyperpolarizability converts local fields into the measured dynamics}

\begin{figure}
\centering
\includegraphics[width=0.90\textwidth]{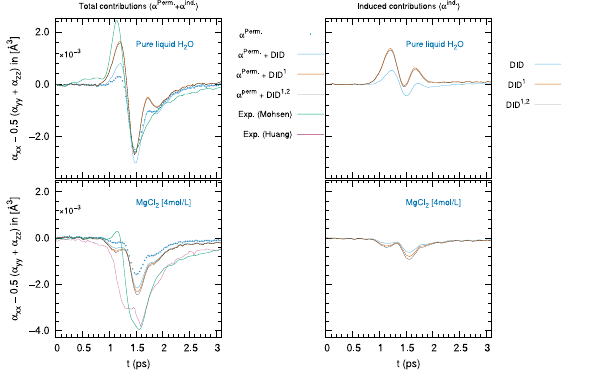}
\caption{\label{fig:did_models}
Transient polarizability anisotropy of pure water and 4~mol/l \mgcl{} computed with the classical DID model and with first- and second-order hyperpolarizability-extended DID models.
The left column shows total responses, while the right column shows induced contributions.
Experimental TKE traces are shown for comparison \cite{HuangHaoyu2020,Hossamenergytransfer,Balos2022,kaliannanzenodo2022}.
The comparison shows that nonlinear molecular response terms are required to convert the time-dependent local field of the liquid into the observed structural-dynamics signal.
}
\end{figure}

Figure~\ref{fig:did_models} compares the classical DID model with the first- and second-order extended DID models.
For neat water, the local intramolecular response alone gives a predominantly negative feature and does not reproduce the experimental bipolar line shape.
The classical DID contribution adds a positive intermolecular component, but the amplitude and balance of the two lobes remain insufficient.
Including the interaction between the first hyperpolarizability and the local electric field strongly enhances the intermolecular contribution, and adding the second hyperpolarizability further refines the response.
The combined permanent and induced response then reproduces the experimentally observed bipolar shape.

The concentrated \mgcl{} solution behaves differently.
The measured signal is dominated by a strong negative birefringence feature \cite{HuangHaoyu2020,Hossamenergytransfer,Balos2022}.
The classical DID model captures the sign and qualitative shape but underestimates the magnitude.
The first and second hyperpolarizability terms systematically increase the amplitude, showing that local-field--hyperpolarizability coupling is not a small correction in strongly ionic environments.
Residual differences to experiment are therefore more naturally attributed to the trajectory, and in particular to whether the underlying force field contains induced polarization.
From the perspective of structural dynamics, this result is important because it shows that the experimental signal cannot be assigned from geometry alone.
The same nuclear configuration carries different optical weight depending on the instantaneous local field and the nonlinear electronic response of the molecules embedded in that field.

\subsection{Induced polarization is required for the concentration-dependent hydration dynamics}

\begin{figure}
\centering
\includegraphics[width=0.82\textwidth]{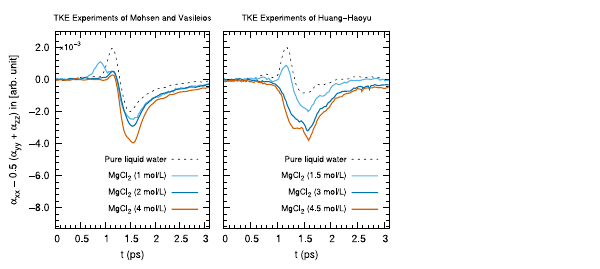}
\vspace{0.4em}
\includegraphics[width=0.82\textwidth]{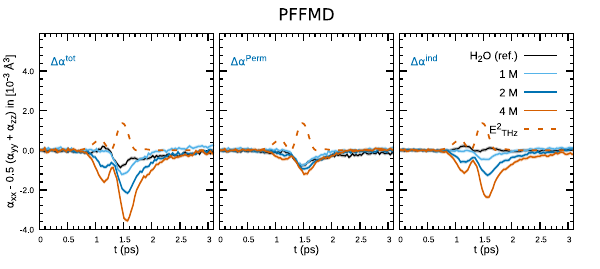}
\vspace{0.4em}
\includegraphics[width=0.82\textwidth]{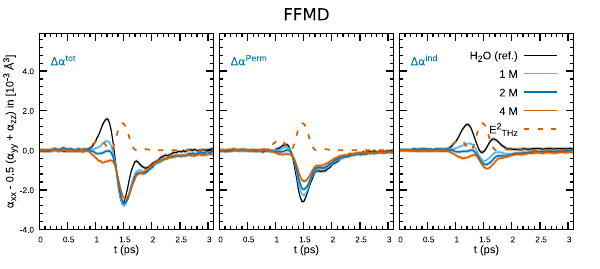}
\caption{\label{fig:mgcl_comparison}
Concentration-dependent transient birefringence and polarizability anisotropy of neat water and aqueous \mgcl{}.
Top: experimental TKE signals.
Middle: PFFMD simulations with the hyperpolarizability-extended DID model.
Bottom: FFMD simulations with the same optical response model.
The polarizable force field reproduces the experimentally observed enhancement with increasing salt concentration, whereas the non-polarizable force field gives an incorrect trend.
The result demonstrates that polarization in the field-driven trajectory is essential for the correct concentration-dependent hydration response.
}
\end{figure}

The concentration dependence of the \mgcl{} response provides a stringent test of the trajectory model.
Figure~\ref{fig:mgcl_comparison} compares experiment, PFFMD, and FFMD for neat water and \mgcl{} solutions.
The PFFMD simulations reproduce the experimentally observed increase in negative birefringence with increasing \mgcl{} concentration.
By contrast, the FFMD simulations fail to reproduce this concentration trend.
This difference demonstrates that induced polarization in the force field is essential for obtaining the correct field-induced anisotropy of concentrated aqueous electrolytes.

The failure of FFMD is not simply a consequence of slower diffusion or orientational relaxation.
As shown below, short-time translational and orientational metrics of the FFMD and PFFMD trajectories are comparable for several observables.
The more important deficiency is the local electronic response: non-polarizable force fields do not adjust the charge distribution to the ion-induced field and therefore do not generate the correct collision-induced polarizability.
The extended DID model can improve the optical response computed from a given trajectory, but it cannot fully compensate for missing induced polarization in the molecular dynamics.
This separates two contributions that are often entangled in liquid spectroscopy: the nuclear structural dynamics of the hydration network and the electronic polarization dynamics that determine how strongly each instantaneous structure is seen by the optical probe.

\subsection{Ion specificity of the transient local-field response}

\begin{figure}
\centering
\includegraphics[width=0.72\textwidth]{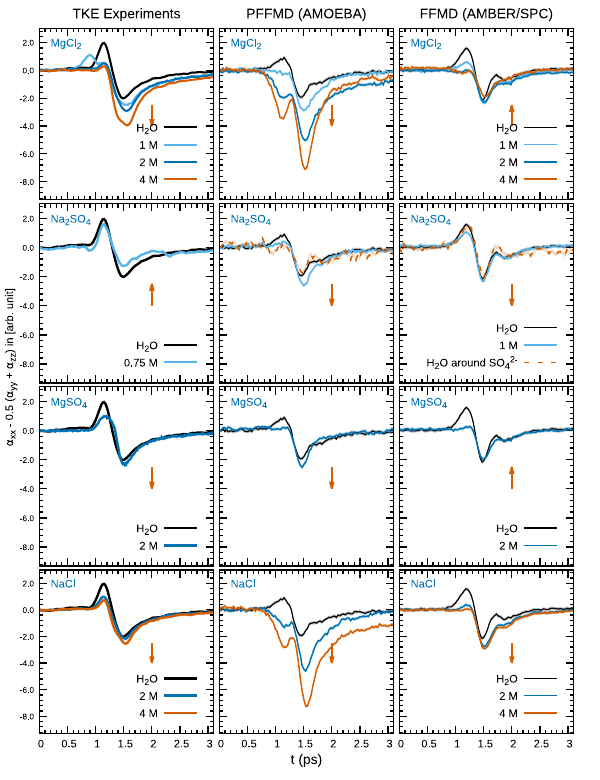}
\caption{\label{fig:salt_trends}
Experimental and simulated transient anisotropy signals for several aqueous salt solutions.
The columns compare experiment, PFFMD, and FFMD, while rows correspond to different salts.
The arrows mark the principal negative feature used to compare concentration and ion-specific trends.
}
\end{figure}

The same analysis was applied to \nasulf{}, \mgso{}, NaCl, and NaF.
Figure~\ref{fig:salt_trends} shows that the polarizable force field gives a broadly consistent description across salts, while the non-polarizable force field often underestimates or misorders the response.
The comparison is especially revealing for salts containing divalent ions.
\mgcl{} and \mgso{} generate strong local fields and enhanced water anisotropy, whereas the response of \nasulf{} and NaCl reflects a different balance between cation hydration, anion polarizability, and the number of waters shared by ion pairs.

The experimental and simulated traces are not expected to agree in every detail because the model separates trajectory generation from optical-response evaluation and relies on gas-phase molecular hyperpolarizabilities.
Nevertheless, the key qualitative result is robust: the sign, amplitude, and concentration dependence of aqueous-electrolyte TKE signals are controlled by the interplay of induced polarization in the trajectory and hyperpolarizability-enhanced local-field response in the optical model.
The ion-specific comparison therefore turns the TKE signal into a sensitive test of the fluctuating electric-field landscape created by different hydration structures.

\subsection{Hydration motifs that dominate the birefringence dynamics}

\begin{figure}
\centering
\includegraphics[width=0.88\textwidth]{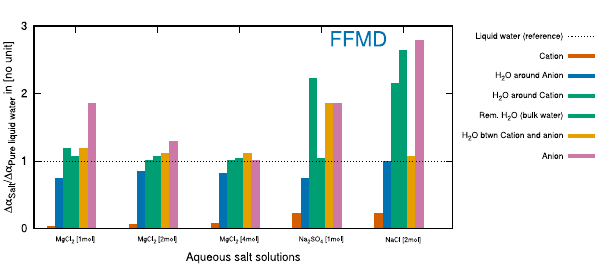}
\vspace{0.3em}
\includegraphics[width=0.88\textwidth]{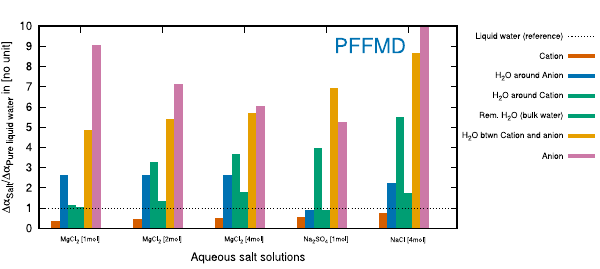}
\caption{\label{fig:species}
Negative peak intensity of the TKE-induced polarizability anisotropy, normalized to pure liquid water, for different molecular sub-ensembles in aqueous salt solutions.
The upper and lower panels show FFMD and PFFMD results, respectively.
The decomposition separates cations, anions, cation-solvating water, anion-solvating water, water bridging cations and anions, and remaining bulk-like water.
The bridging-water component provides the strongest molecular assignment of the enhanced birefringence in concentrated electrolytes.
}
\end{figure}

To identify which molecular environments dominate the salt-induced response, the computed anisotropy was decomposed into cations, anions, water in the cation solvation shell, water in the anion solvation shell, water bridging cations and anions, and remaining bulk-like water.
The peak amplitudes normalized to neat water are shown in Fig.~\ref{fig:species}.
Both FFMD and PFFMD show that cations have only a small direct polarizability anisotropy.
Anions, by contrast, are strongly polarized by the applied field and contribute a large anisotropic response.
This is consistent with their larger electronic polarizabilities relative to small cations \cite{amoebapolarisability2,polarisabilityCAnMg}.

The dominant contribution to the overall response, however, often comes from water.
Water molecules located between cations and anions display particularly large anisotropies, and their population increases in concentrated salt solutions.
These bridging waters are exposed to strong and anisotropic local fields from both ionic partners.
As a result, their hyperpolarizability-enhanced DID response is larger than that of bulk-like water or water coordinated to only one type of ion.
The species decomposition therefore assigns the strong concentration-dependent TKE response primarily to interionic hydration environments rather than to the bare ions alone.
This is the main structural assignment of the work: the enhanced negative birefringence is a signature of transiently shared hydration waters and their local-field environment, not merely of the presence of more ions in solution.

\subsection{Coupling between structural relaxation and optical anisotropy}

\begin{figure}
\centering
\includegraphics[width=0.90\textwidth]{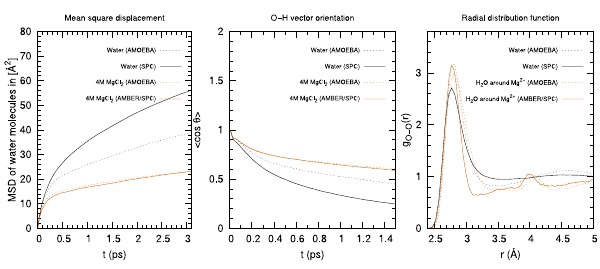}
\vspace{0.3em}
\includegraphics[width=0.44\textwidth]{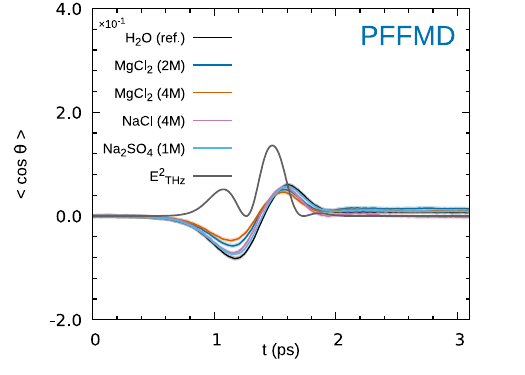}
\includegraphics[width=0.44\textwidth]{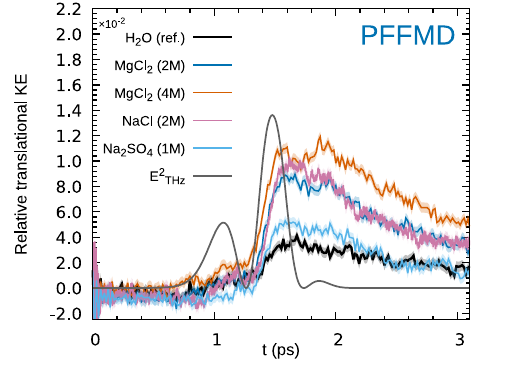}\\[-0.1em]
\includegraphics[width=0.44\textwidth]{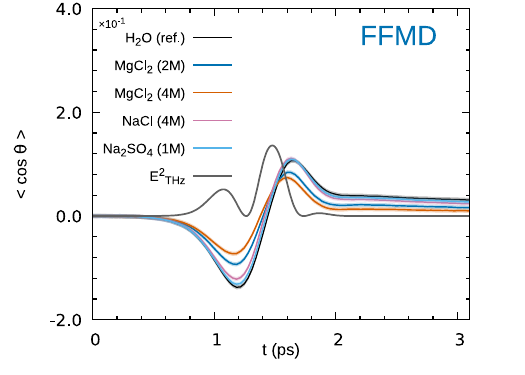}
\includegraphics[width=0.44\textwidth]{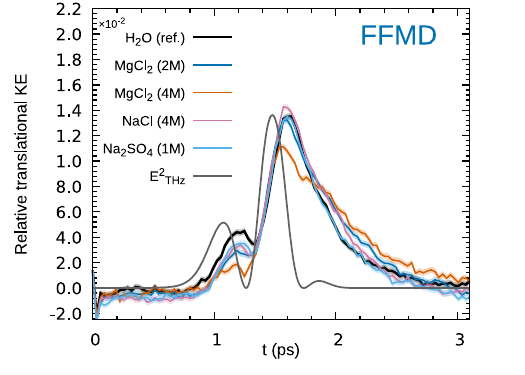}
\caption{\label{fig:dynamics}
Structural and dynamical observables for pure water and \mgcl{} solutions.
The top panels show mean-square displacements, O--H orientational relaxation, and O--O radial distribution functions.
The lower panels show O--H orientational and translational kinetic-energy relaxation for PFFMD and FFMD trajectories.
These observables connect the transient optical anisotropy to the relaxation of the hydrogen-bond and ion-hydration network.
}
\end{figure}

Figure~\ref{fig:dynamics} compares short-time translational and orientational dynamics of neat water and 4~mol/l \mgcl{}.
The salt solution modifies the mean-square displacement, the O--H orientational relaxation, and the O--O radial distribution function, as expected for a strongly hydrated electrolyte.
The RDF is particularly relevant because the DID contribution depends on intermolecular distances through the dipole--dipole tensor and damping functions.
Small changes in the first and second water shells can therefore have a disproportionate effect on the computed optical response.

The kinetic-energy traces further show that the experimentally relevant TKE time window overlaps with the relaxation of excess translational energy.
This supports the assignment of the negative tail in water-rich systems to collective hydrogen-bond and translational relaxation, while the enhanced response in concentrated salts reflects additional local-field contributions from ion hydration and water molecules shared by ion pairs.
The optical anisotropy is therefore a coupled structural-electronic observable: its time dependence follows the relaxation of the hydrogen-bond and ion-hydration network, while its amplitude is amplified by nonlinear local-field response.

\section{Conclusions}
\label{sec:conclusions}

We have developed a molecular simulation framework that connects terahertz-induced transient birefringence to the structural and local-field dynamics of water and aqueous electrolytes.
The approach combines terahertz-driven NEMD trajectories with a hyperpolarizability-enhanced DID optical readout in which first and second molecular hyperpolarizabilities are coupled to the instantaneous local electric field.
This coupling is essential for reproducing the bipolar response of neat water and for recovering the experimentally observed amplitude enhancement in concentrated salt solutions.

For aqueous \mgcl{}, the polarizable trajectories reproduce the concentration-dependent enhancement of the negative birefringence, whereas non-polarizable trajectories yield the wrong trend.
The comparison shows that induced polarization is required twice: first in the field-driven molecular dynamics that generates the fluctuating hydration structure, and again in the optical response model that weights each instantaneous structure by its local field.
The species-resolved decomposition assigns the strongest electrolyte response to anions and, in particular, to water molecules bridging cations and anions.
The dominant structural motif is therefore not a bare ion, but an interionic hydration environment whose local field strongly amplifies the nonlinear optical response.

The framework is computationally efficient enough for statistically converged simulations of nonlinear optical signals in complex liquids, while retaining a direct molecular assignment of the measured response.
It should therefore be useful for interpreting TKE and related ultrafast birefringence measurements of electrolytes, hydrogen-bonded liquids, interfaces, and confined aqueous systems where the observable is governed by coupled structural relaxation and local-field dynamics.

\section*{Supplementary Material}

See the supplementary material for the detailed multipolar derivation of the differential polarizability used in the hyperpolarizability-enhanced DID model, the induced-dipole self-consistency equations, validation calculations for the optical-response code, and the terahertz driving pulse used in the simulations.

\begin{acknowledgments}
The authors gratefully acknowledge the computing time made available to them on the high-performance computer Noctua at the NHR Center Paderborn Center for Parallel Computing (PC2).
This center is jointly supported by the Federal Ministry of Research, Technology and Space and the state governments participating in the National High-Performance Computing (NHR) joint funding program (www.nhr-verein.de/en/our-partners).
Part of the research was funded by the DFG (project numbers 398046241, 417590517/CRC1415 and 519869949).
\end{acknowledgments}

\section*{Author Declarations}

\subsection*{Conflict of Interest}
The authors have no conflicts to disclose.

\subsection*{Author Contributions}
Naveen Kumar Kaliannan: Investigation, methodology, software, visualization, writing -- original draft.
Thomas D. K{\"u}hne: Conceptualization, formal analysis, funding acquisition, project administration, supervision, writing -- review and editing.
Hossam Elgabarty: Conceptualization, formal analysis, methodology, validation, writing -- review and editing.

\section*{Data Availability Statement}
The input files, trajectory-analysis scripts, source-code archive, and source data associated with this work are available from Zenodo at \url{https://doi.org/10.5281/zenodo.6514905} \cite{kaliannanzenodo2022}.
The polarizability-analysis code is available at \url{https://github.com/NaveenKaliannan/Dipole-Polarisability} \cite{kaliannandipol2022}.
Additional data that support the findings of this study are available from the corresponding author upon reasonable request.

\bibliographystyle{aipnum4-1}
\bibliography{references}

\end{document}